# Leveraging local h-index to identify and rank influential spreaders in networks

Qiang Liu[a,*], Yuxiao Zhu[b], Yan Jia[a,c], Lu Deng[a], Bin Zhou[a,c], Junxing Zhu[a], Peng Zou[a]

*Corresponding author:liuqiang1981@nudt.edu.cn

[a]*College of Computer, National University of Defense Technology, Changsha 410073, China*

[b]*School of Management, Guang Dong University of Technology, Guangzou,510006,China*

[c]*State Key Laboratory of High Performance Computing, National University of Defense Technology, Changsha 410073, China*

**Abstract:** Identifying influential nodes in complex networks has received increasing attention for its great theoretical and practical applications in many fields. Some classical methods, such as degree centrality, betweenness centrality, closeness centrality, and coreness centrality, were reported to have some limitations in detecting influential nodes. Recently, the famous h-index was introduced to the network world to evaluate the spreading ability of the nodes. However, this method always assigns too many nodes with the same value, which leads to a resolution limit problem in distinguishing the real influences of these nodes. In this paper, we propose a local h-index centrality (LH-index) method to identify and rank influential nodes in networks. The LH-index method simultaneously takes into account of h-index values of the node itself and its neighbors, which is based on the idea that a node connecting to more influential nodes will also be influential. Experimental analysis on stochastic Susceptible-Infected-Recovered (SIR) model and several networks demonstrates the effectivity of the LH-index method in identifying influential nodes in networks.

## 1. Introduction

Identifying influential nodes in networks has become a hot topic in recent years for its wide applications in many fields, such as social network analysis [1-3], viral marketing [4], epidemic spreading and controlling [5, 6]. By finding influential nodes, we can have a better understanding of the characteristics of the network structure and function so as to fulfill various applications being closely related to human lives. Meanwhile, some further applications based on influential nodes identification also receive increasing attention from scholars. For instance, some have utilized influence propagation method to find densely connected parts of the networks [7, 8], while some others try to find laws of information transmission in networks by utilizing influential nodes [9, 10]. Eventually, all these methods contribute greatly to requirements of new ways for identifying influential nodes in networks. Therefore, how to effectively identify influential nodes in networks has become an urgent problem.

Traditional centrality measures, such as degree centrality [11], betweenness centrality [12], closeness centrality [13] and coreness centrality [14], have been adopted to evaluate the influence of the nodes. Given one node, degree centrality [11] measures its influence by counting the number of its connected neighbors. However, the degree centrality neglects the location and structure information of a node, which may not fairly reflect the real influence of the node. For example, one high degree node may be located in the periphery of the network, then it can only affect very few other nodes. Both betweenness centrality [12] and closeness centrality [13] are global centrality methods, and they measure the influence by computing the number of the shortest paths in the whole network, so these two methods are not fit for large-scale networks due to their high computational complexity. Coreness centrality [14] measures the influence by k-core decomposition, the node with higher coreness value means that its location is more central in the network. Note that, the computing of coreness needs global topological information of the whole network.

In recent years, some new centrality measures have been put forward. Lü et al. [15] devised an adaptive and parameter-free LeaderRank algorithm to quantify influential nodes in social networks, which achieves good performance for directed networks but is weak for undirected networks. Chen et al. [16] proposed a semi-local centrality measure which considers both the nearest and the next nearest neighbors' degree information of a node. However, the degree information of a node sometimes can not reflect its real influence fairly, because the influences of different neighbors always vary greatly. Whereafter, some core-based methods were proposed. Bae et al. [17] studied the neighbor's coreness centrality of a node for identifying influential nodes in networks. Ma et al. [18] proposed the neighbor's gravity centrality to detect influential nodes in networks. Liu et al. [19] generalized the neighbors' centrality methods [16, 17] in networks, and made detailed analyses of the neighbor's degree and coreness centrality. Malliaros et al. [20] capitalized on the properties of the K-truss decomposition to locate influential nodes, which can filter out less important information and detect influential nodes. Basu et al. [21] proposed a group density method based on core analysis to identify influential nodes in weighted networks, this method was inspired by the idea of game theoretic concept of voting mechanism. However, the core-based method always needs global topological information of the network, thus methods of this kind always own high computational cost, which hinders their practical applications in large networks. Besides, recent research [22] pointed out that nodes with the highest coreness may not be the most influential spreaders due to the existence of the core-like groups. Later, Liu et al. [23] found that the core-like groups can be attributed to the existence of redundant links, then they proposed a new method to improve the accuracy of the K-core decomposition by removing these redundant links. Nonetheless, all the centrality measures that mentioned above are sensitive to the small variation of nodes' degree information. For instance, if the collected dataset missed some connection information of few nodes, then the influences of these nodes, which are measured by degree centrality or coreness centrality, will be affected directly. The h-index, proposed by Hirsch in 2005 [24], is a classical metric to measure both the productivity and citation impact of the publications of a scholar. Recently, some scholars introduced the h-index to measure the spreading influence of the nodes in networks. For example, Korn et al. [25] illustrated that the h-index can make a well-balanced mix of traditional centrality measures. Lü et al. [26, 27] also demonstrated the effectivity of the h-index in evaluating the influence of the nodes in many real world networks. However, we find that the h-index method always assigns the same value to many different nodes, which makes it difficult to distinguish the real influences of these nodes. So it's necessary to design one better ranking method which can improve its resolution power in identifying influential nodes.

Inspired by the classical h-index, we put forward one novel method to better qualify the spreading ability of the nodes in this paper. The new proposed local h-index (LH-index) method simultaneously considers two factors: the h-index value of the node itself and the h-index values of its neighbors. On one hand, the h-index of one node indicates its direct influence. On the other hand, the h-index values of its neighbors indicate its two-hop indirect influence. That is to say, the node which is surrounded by much more high h-index neighbors is more influential. We adopt the epidemic spreading process to evaluate the performance of LH-index in identifying influential nodes, The experimental results demonstrate that LH-index outperforms many traditional methods in many networks.

The rest of this paper is organized as follows. In Section 2, we will briefly review some centrality measures and present the definition of the new LH-index. In Section 3, we will introduce the network datasets, the spreading model and the evaluation method. In Section 4, we will present our experimental results. Finally, conclusions will be given in Section 5.

## 2. Methods

In this paper, we consider four classical centrality measures, which are degree centrality, betweenness centrality, closeness centrality and coreness centrality. Consider an undirected network $G=(V,E)$, where $V$ is the vertex set, and $E$ is the edge set. The degree centrality (DC) [11] of node $i$ is defined as the number of $i$'s nearest neighbors. The betweenness centrality (BC) [12] of node $i$ is defined as the fraction of the number of shortest paths that travel through the node $i$ to all the shortest paths. The closeness centrality (CC) [13] measures how close a node is to all the other nodes, which is defined as the reciprocal of the sum of the shortest distances to all other nodes.

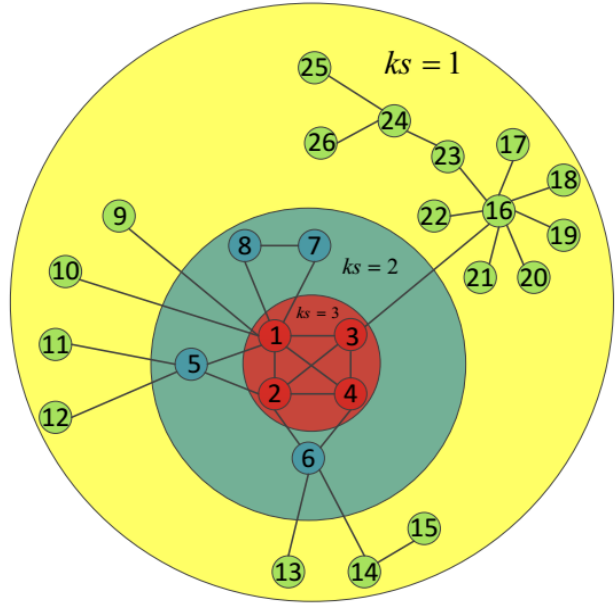

**Fig. 1.** A schematic representation of a network under the k-core decomposition presented in Ref [14]

K-core centrality (KC) [14] measures a node's influence based on its location. K-core is the maximal connected subgraph of $G$, in which all nodes' degree values are equal to or larger than $k$. The k-core decomposition method [28] classifies all nodes into k-shells by removing nodes iteratively as follows: First, all the nodes with degree $k=1$ will be removed, which will lead to the degree reduction of the remaining nodes. These removed nodes form the 1-shell and are assigned with the coreness value $ks=1$. Then we continually remove all the nodes with degree $k=2$ until there are only nodes with degree $k>2$, these removed nodes form the 2-shell and are assigned with the coreness value $ks=2$ This procedure is repeated until all nodes are removed and assigned to one of the k-shells. It's obvious that the node with a larger cornness value is always located in a much more central position of the network.

In order to describe the k-core decomposition clearly, we take one simple network [14,17,27] for example. In Fig. 1, node 1 and node 16 own the same degree value with $k=8$, so the degree centrality cannot distinguish which one is much more influential effectively. By the k-core decomposition of the network, we can see that node 16 lies in the periphery of the network with coreness value $ks=1$, while node 1 is located in the innermost of the network with the largest coreness value $ks=3$. So a node with a larger coreness value lies in a much more central location in the network, and owns higher influence. However, we can see from Fig. 1 that there are many nodes with the same coreness value while these nodes have different connection patterns with other nodes, which can demonstrate the weakness of the coreness centrality. Besides, we can also find that the KC method always require the global topological information of the network to calculate the coreness value of a node.

Recently, the h-index [24] (also known as Hirsch index) is introduced to identify influential spreaders [25-27] in networks. The h-index of node $i$ is defined as the largest value $h$ such that $i$ has at least $h$ neighbors for each with a degree no less than $h$ [25]. For example, in Fig. 1, we can see that node 1 has 4 neighbors for each with a degree at least 4, so the h-index of node 1 is 4. However, for node 16, although its degree value is equal to node 1, its h-index value is 2. Then, we can see that the h-index can evaluate a node's influence according to the number of high-quality influential neighbors, and does not need the global topological information of network.

Besides, the h-index of a node is not sensitive to the small variation of the degree information of itself and neighbors, which is beneficial to evaluate a node's spreading ability in networks with information of few links missing or being incorrect. Note that, it is really difficult to ensure that the collected datasets are complete without any mistake of linkage information in many cases. For instance, in Fig. 1, if we add a link between node 1 and node 11, the h-index value of node 1 will not be affected ; similarly, if the links between node 1 and node 7,8,9,10 miss, the h-index of node 1 will still keep the same. That is to say, for a node , if the degree values of some neighbors changed, the h-index value of this node may not be affected. That is because h-index only takes into account the number of neighbors with high-quality degree information, so the small variation of few neighbors may not induce the variation of the node's overall spreading ability. As far as other centrality measures are concerned, such as degree centrality and betweenness centrality, the missing or incorrect linkage information of few edges will affect the ranking results obviously. However, h-index also has its weakness. The h-index always assigns same value for many nodes with different influences, which leads to a resolution limit problem in distinguishing the real spreading ablities of these nodes. For example, Table 1 presents the centrality values of nodes in Fig. 1, we can find that there are only three different h-index values for all 26 nodes, which cannot distinguish the real spreading influences of these nodes effectively.

**Table 1** The centrality values of degree centrality (DC), betweenness centrality (BC), closeness centrality (CC), coreness centrality (KC), H-index centrality and local h-index (LH-index) centrality, which correspond to nodes in Fig. 1.

| Node | DC | BC | CC | KC | H-index | LH-index |
|---|---|---|---|---|---|---|
| 1 | 8 | 108.500 | 0.410 | 3 | 4 | 24 |
| 2 | 5 | 65.500 | 0.410 | 3 | 4 | 20 |
| 3 | 4 | 154 | 0.438 | 3 | 4 | 18 |
| 4 | 4 | 34 | 0.391 | 3 | 4 | 18 |
| 5 | 4 | 47 | 0.329 | 2 | 2 | 12 |
| 6 | 4 | 68 | 0.321 | 2 | 2 | 12 |
| 7 | 2 | 0 | 0.298 | 2 | 2 | 8 |
| 8 | 2 | 0 | 0.298 | 2 | 2 | 8 |
| 9 | 1 | 0 | 0.294 | 1 | 1 | 5 |
| 10 | 1 | 0 | 0.294 | 1 | 1 | 5 |
| 11 | 1 | 0 | 0.250 | 1 | 1 | 3 |
| 12 | 1 | 0 | 0.250 | 1 | 1 | 3 |
| 13 | 1 | 0 | 0.245 | 1 | 1 | 3 |
| 14 | 2 | 24 | 0.250 | 1 | 1 | 4 |
| 15 | 1 | 0 | 0.202 | 1 | 1 | 2 |
| 16 | 8 | 189 | 0.410 | 1 | 2 | 14 |
| 17 | 1 | 0 | 0.294 | 1 | 1 | 3 |
| 18 | 1 | 0 | 0.294 | 1 | 1 | 3 |
| 19 | 1 | 0 | 0.294 | 1 | 1 | 3 |
| 20 | 1 | 0 | 0.294 | 1 | 1 | 3 |
| 21 | 1 | 0 | 0.294 | 1 | 1 | 3 |
| 22 | 1 | 0 | 0.294 | 1 | 1 | 3 |
| 23 | 2 | 66 | 0.316 | 1 | 2 | 5 |
| 24 | 3 | 47 | 0.253 | 1 | 1 | 5 |
| 25 | 1 | 0 | 0.203 | 1 | 1 | 2 |
| 26 | 1 | 0 | 0.203 | 1 | 1 | 2 |

Considering the weakness of the h-index, we put forward a modified method. The new proposed method, named local h-index (LH-index) centrality , can not only inherit all advantages of h-index method but also resolve the weakness of it. The LH-index value of node $i$ is defined as follows:

$$LH_{index}(i) = h_{index}(i) + \sum_{v \in N(i)} h_{index}(v), \quad (1)$$

where $N(i)$ is the set of neighbors of node $i$, $h_{index}(i)$ represents the h-index value of node $i$, and $LH_{index}(i)$ is the LH-index value of node $i$. The LH-index simultaneously considers the spreading power of the node itself and its neighbors. The LH-index is composed of two parts: the first part is the h-index of a node, which evaluates its spreading ability according to the number of high-quality neighbor nodes. Take the citation network as an example, the h-index evaluates a scholar's influence according to both the number of the scholar's publications and the number of citations, named high-quality publications. However, for a network, the resolution power of the h-index is limited due to the fact that it assigns too many nodes with the same h-index value. Therefore, we also take the neighbors' h-index information into account. Then, the second part sums the h-index values of neighbor nodes together, which means that if a node has more neighbors with higher h-index values, it will be more influential. Here we take node 3 in Fig. 1 as example to describe how to calculate the LH-index value. The h-index value of node 3 is 4. From Fig. 1, we can see that there are four neighbor nodes of node 3, which are node 1, 2, 4 and 16, and their h-index values are 4, 4, 4, and 2, respectively. So the second part of LH-index for node 3 is 4+4+4+2=14. Eventually, the LH-index value of node 3 is 4+14=18. By the same way, we can calculate the LH-index values of all nodes in Fig. 1, which were shown in Table 1.

Next, let's consider the computational complexity of LH-index centrality. Given one network with $n$ nodes, the time complexity of calculating h-index of each node $i$ is $O(\langle k \rangle)$, where $\langle k \rangle$ is the average degree of the network. Then, the time complexity for computing h-index values of all nodes is $O(n\langle k \rangle)$. Note that, for node $i$, the LH-index needs traversing its neighbor nodes within one step, so the total computational complexity for LH-index is $O(n\langle k \rangle)$. Obviously, the new LH-index is much more efficient compared to other traditional methods, such as betweenness and closeness centrality.

## 3. Experimental setup

### 3.1 Dataset

To test the effectivity of the LH-index method, we apply it to both real world networks and simulated networks. The real world networks include (i) USair: the network of the US air transportation [29]. (ii)Blogs: the network of the communication relationships between owners of blogs on the MSN (Windows live) spaces website [16]. (iii)Email: the network of e-mail interchanges between members of the University Rovira i Virgili(Tarragona) [30]. (iv) Powergrid: the network of the power grid of the Western States in the United States of America [31]. The basic structural properties of these four networks are shown in Table 2. The simulated networks contain networks generated by two kinds of network models: the Barabasi-Albert (BA) network model [32], and the Lancichinetti-Fortunatio-Radicchi (LFR) network model [33]. In this paper, all of these real world networks and simulated networks are treated as undirected and unweighted.

**Table 2** The basic structural properties of four real world networks. $n$ and $m$ are the number of nodes and edges, respectively. $\beta_c$ is the epidemic threshold ($\beta_c = \langle k \rangle / \langle k^2 \rangle$ [34]). $\langle k \rangle$ and $k_{max}$ are the average degree and maximum degree. $C$ is the clustering coefficient. $\langle d \rangle$ is the average shortest distance. $r$ is the assortativity coefficient.

| Network   | $n$  | $m$  | $\beta_c$ | $\langle k \rangle$ | $k_{max}$ | $C$    | $\langle d \rangle$ | $r$    |
|-----------|------|------|-----------|---------------------|-----------|--------|---------------------|--------|
| USair     | 332  | 2126 | 0.0225    | 12.81               | 139       | 0.6252 | 2.738               | -0.208 |
| Blogs     | 3982 | 6803 | 0.0725    | 3.42                | 189       | 0.2838 | 6.252               | -0.133 |
| Email     | 1133 | 5451 | 0.0535    | 9.62                | 71        | 0.2202 | 3.606               | 0.078  |
| Powergrid | 4941 | 6594 | 0.2583    | 2.67                | 19        | 0.0801 | 18.989              | 0.003  |

### 3.2 The stochastic SIR epidemic model

Currently, the Susceptible-Infected-Recovered (SIR) [35] is widely used in illustrating and simulating the spreading process of epidemic or information [14, 36-38]. In this paper, we adopt the stochastic Susceptible-Infected-Recovered (SIR)[39] model to evaluate the real spreading ability of the ranking nodes. In the stochastic SIR model, an individual can be in one of three possible states: susceptible(S), infected (I), or recovered (R). In the susceptible state, an individual does not adopt the information. In the infected state, an individual adopts the information and tries to transmit the information to his/her selected neighbors. In the recovered state, an individual lose interest in the information and will not transmit the information further. In our experiments, we set one single node in the infected state and other nodes in the susceptible state initially. Then, the infected node tries to infect its susceptible neighbors with the transmission rate $\beta$. At each time step, each infected node will choose infection event or recover event according to the Gillespie algorithm [40]. The recovery rate is set to 1, which means that if a node selects recovery event, it will enter into the recovery state. This process repeats until there are no infected nodes in the network. Obviously, if we select node $i$ as the initial single seed in the stochastic SIR model, then the final number of recovered nodes can denote the real spreading ability of node $i$. Through this, we can get the ranking of all the nodes in the network and then identify the influential ones.

Besides, the outbreak of the spreading process is related to the epidemic threshold [41]. If the transmission rate is much smaller than the threshold, the spreading process will end eventually. On the contrary, the spreading process will disseminate a large fraction of the network if the transmission rate is much larger than the epidemic threshold. Therefore, it is reasonable to choose the neighborhood of the epidemic threshold to evaluate the effectivity of a centrality measure. That is to say, too smaller value of the transmission rate will lead to the finite size of spreading process, while too larger value of the transmission rate will result in large-scale diffusion of the network, which is independent of the single spreader.

### 3.3 Evaluation method

Here we employ the Kendall's tau correlation coefficient [42] $\tau$ to measure the performance of various centrality measures. Kendall's tau correlation coefficient is a measure of rank correlation: the similarity of the orderings of the data when ranked by each measure. Intuitively, the Kendall correlation between two variables will be high when two variables have a similar rank, and low when two variables have a dissimilar rank. Given one certain centrality measure, by calculating the Kendall's tau correlation coefficient between the ranking list of this measure and the ranking list obtained by the stochastic SIR model, we can get the performance of this measure easily. If the $\tau$ is more close to 1, it means that the ranking list acquired by this kind of centrality method is more consistent with the nodes' real spreading ability, which can demonstrate the better performance of this method.

Let $(x_1,y_1)$, $(x_2,y_2)$, …, $(x_n,y_n)$ be a set of joint ranks from two ranking lists $X$ and $Y$ respectively. In this paper, $X$ can denote the ranking result of one centrality measure, and $Y$ can represent the real ranking result of the number of infected nodes. Then, any pair of ranks $(x_i,y_i)$ and $(x_j,y_j)$ are said to be concordant if both $x_i > x_j$ and $y_i > y_j$ or if both $x_i < x_j$ and $y_i < y_j$. That is to say, the ranking order of $x_i$ and $x_j$ in $X$ is consistent with the ranking order of $y_i$ and $y_j$ in $Y$. Besides, $(x_i,y_i)$ and $(x_j,y_j)$ are said to be discordant if both $x_i > x_j$ and $y_i < y_j$ or if both $x_i < x_j$ and $y_i > y_j$. In other words, the ranking order of $x_i$ and $x_j$ in $X$ is contrary to the ranking order of $y_i$ and $y_j$ in $Y$. If $x_i = x_j$ or $y_i = y_j$, the pair is neither concordant or discordant. The Kendall's tau correlation coefficient $\tau$ is defined as[42]:

$$\tau = \frac{N_c - N_d}{\frac{1}{2}n(n-1)}, \quad (2)$$

where $N_c$ is the number of concordant pairs, $N_d$ is the number of discordant pairs, and the denominator $\frac{1}{2}n(n-1)$ is the total number of pair combinations. If the ranking orders of two ranking lists are identical, which means $N_c = \frac{1}{2}n(n-1)$ and $N_d = 0$, then $\tau = 1$. On the contrary, if the ranking order of one ranking list is the reverse of the other, which means $N_c = 0$ and $N_d = \frac{1}{2}n(n-1)$, then $\tau = -1$. Besides, if the ranking orders of two ranking lists are independent, then $\tau = 0$. In this paper, $\tau = 1$ indicates the ranking list generated by a centrality measure is the same as the ranking list generated by spreading process, and $\tau = -1$ means these two ranking lists are reversed.

## 4  Experimental results

In Section 3.2, we have illustrated the importance of epidemic threshold for spreading process. For the four real world networks listed in Table 2, the epidemic thresholds of the USair network, Blog network, and Email network are 0.0225, 0.0725, and 0.0535, respectively, so the variation range of transmission rate for the three networks is set to [0.01,0.10]. For the Powergrid network, the epidemic threshold is 0.2583, and the variation range of transmission rate is set to [0.21, 0.30]. The Section 4.1, 4.2 and 4.3 demonstrate the performance of LH-index and other centrality measures in these four real world networks. In Section 4.4, we clarify the performance of these centrality measures in simulated networks. As far as these simulated networks are concerned, their epidemic thresholds are determined by the parameters utilized to generate them.

### 4.1 The performance of different centrality measures under different transmission rate $\beta$

In order to compare the performance of DC, BC, CC, KC, H-index and LH-index, we calculate the Kendall's tau values for each method under different transmission rate $\beta$, which are presented in Fig.2.

We can see from Fig.2 that the $\tau$ values of these six methods vary with the increasing of $\beta$. DC performs better when $\beta$ is smaller, and with the increasing of $\beta$, the $\tau$ value of DC begins to decrease first and then grow or drop off slowly based on different network structures. For instance, in Fig. 2(a) and Fig. 2(c), DC is better than other methods when $\beta$ is smaller than 0.01. That is because smaller $\beta$ leads to the limited spreading area, which is confined to the neighborhood of the initially infected node. So a node with a larger degree will have more chance to infect its neighbor nodes. Then, when $\beta$ increases gradually and approximates to the epidemic threshold $\beta_c$, DC cannot infect more nodes due to the fact that $\beta$ is still smaller than $\beta_c$, which eventually leads to the decreasing of the $\tau$ value. Only when $\beta$ is larger than $\beta_c$, the $\tau$ value of DC begins to increase, due to the improved spreading ability of the infected node, which will have more chance to infect further neighbors. In Fig. 2(b) and Fig. 2(d), the $\tau$ value of DC decreases slowly when $\beta$ is larger than $\beta_c$, that is because Blogs and Powergrid are sparse networks with low average degree, and the increasing of transmission rate only can infect a limited number of nodes.

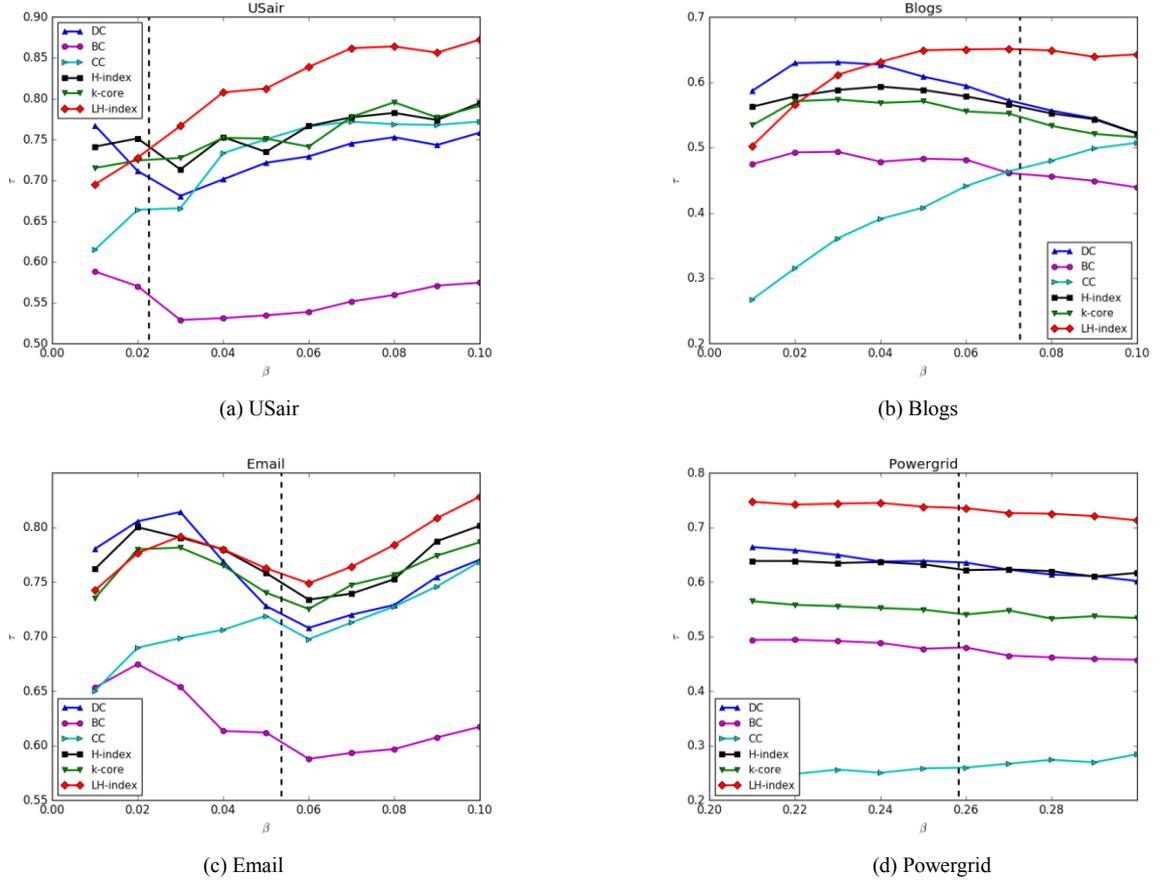

(a) USair
(b) Blogs
(c) Email
(d) Powergrid

**Fig. 2.** The $\tau$ values of DC, BC, CC, KC, H-index and LH-index are obtained by averaging over 100 implementations of the stochastic SIR model under different transmission rate $\beta$ in USair, Blogs, Email and Powergrid. The vertical dashed line for each plot corresponds to the epidemic threshold $\beta_c$, which is given in Table 2.

As we can see from Fig. 2, LH-index achieves the best performance in a large range of $\beta$ in four real world networks. LH-index gets higher $\tau$ values when $\beta$ is close to $\beta_c$, and it's performance is also superior to other measures with the growth of $\beta$. Besides, when $\beta$ is smaller than $\beta_c$, H-index method is better than LH-index, but worse than DC, that is because H-index only considers the information of the influential neighbor nodes directly around the infected node, while LH-index takes into account of farther information of the infected node. Therefore, only when $\beta$ is close to or larger than $\beta_c$, LH-index achieves better performance compared to H-index and DC. Note that, even though H-index is not as good as LH-index, it still has advantages compared with other tranditional measures in most cases. Besides, BC cannot get better results in all these networks. That is because the spreading of information does not always go along with the shortest path between nodes. For example, in a social network, some information concerning a friend may not be directly from this friend but from a mutual friend. Therefore, the performance of BC method based on counting the number of shortest paths between nodes is limited.

**4.2 The correlation between centrality value and spreading ability**

In order to investigate the the effectivity of various centrality measures, the stochastic SIR model is utilized to simulate the spreading process. Then we adopt the Kendall's tau correlation coefficient to measure the performance of various centrality measures. Specifically, given one certain centrality measure, the Kendall's tau correlation coefficient between the ranking list of this measure and the ranking list obtained by SIR model can reflect the performance of this measure. If one centrality measure can reflect the real spreading ability of nodes effectively, the similarity of two rankings ($\tau$) will be high. Take the Email and Powergrid network as examples, the Kendall's tau correlation coefficients between different centrality measures and their real spreading ability are shown in Fig. 3.

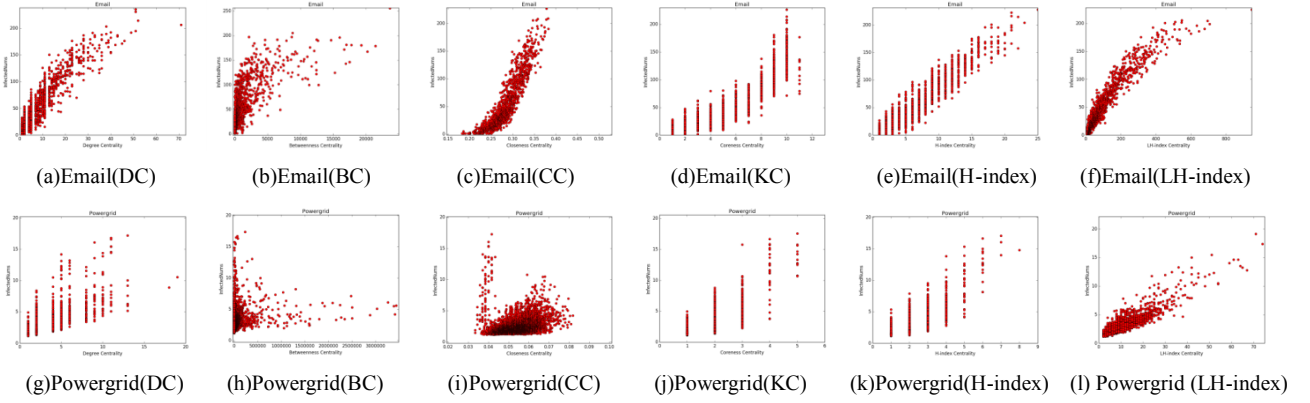

(a)Email(DC)  (b)Email(BC)  (c)Email(CC)  (d)Email(KC)  (e)Email(H-index)  (f)Email(LH-index)

(g)Powergrid(DC)  (h)Powergrid(BC)  (i)Powergrid(CC)  (j)Powergrid(KC)  (k)Powergrid(H-index)  (l) Powergrid (LH-index)

**Fig. 3.** The correlation between node's infected number and its value of each centrality measure. For each network dataset, there are six graphs representing six centrality measures, which are degree centrality (DC), betweenness centrality (BC), closeness centrality (CC), coreness centrality (KC), H-index centrality and local h-index centrality (LH-index), respectively. The transmission rate is 0.1 for Email and 0.3 for Powergrid according to their epidemic thresholds. The values are obtained by averaging over 100 implementations of the stochastic SIR model.

As can be seen from Fig. 3, for the Email network, all these measures except betweenness centrality (BC) present a better correlation with the number of infected nodes, and the corresponding results are plotted in Fig. 2(c). Besides, by observing Fig. 3(d) and Fig. 3(e), we can find that KC and H-index assign the same rank value to many nodes which have different spreading ability, this can demonstrate the limits of KC and H-index methods discussed previously in Section 2. For the Powergrid network, the performance of DC and CC is not as good as the Email network. That is because Powergrid is sparser than Email, which means there are less links. Therefore, too many nodes are assigned with the same degree value. Obviously, this condition is similar for KC and H-index, which leads to the declining of the performance of DC, such as Fig. 3(g). Besides, CC evaluates a node's influence according to its shortest distances to all the other nodes, but the average shortest distance for the Powergrid is 18.989 (Table.1), which directly limits the performance of CC, such as Fig. 3(i). Meanwhile, this condition is in accordance with Fig. 2(d) of Section 4.1, because the $\tau$ values of CC are always smaller than 0.3, then CC achieves worst performance in this network. However, LH-index is still competent in better representing the nodes' real spreading ability compared to others, such as Fig. 3(l).

From what has been analyzed above, we can see that the ranking list of LH-index is more consistent with the nodes' real influence, which can explicate its better performance presented in Fig. 2.

### 4.3 The correlations between LH-index and other centrality measures

In this part, we present the correlations between LH-index and other centrality measures in four real world networks in Table 3. At the same time, we still take the Email and Powergrid as examples, and present their correlation graphs in Fig. 4, which aims to check whether the proposed LH-index method is correlated with other centrality measures. In Fig. 4, each point represents a node in the network, and its color indicates the infected number of nodes, which can be evaluated approximately according to the color bar on the right side of the plot. Take Fig. 4(a) as an example, it presents the correlation between LH-index and degree centrality (DC) in Email network. The node with blue color shows the infected number of nodes is around 25, and the node with red color means the infected number of nodes is about 200. With the increasing of LH-index values of nodes, the DC values of them also grow, meanwhile, the infected number of nodes ranges from blue to red, so we can make a conclusion that LH-index is positively correlated with DC in this network. Besides, by adding different colors of nodes, we can easily distinguish the nodes' real spreading ability and correctly estimate the correlations between LH-index and other centrality measures.

**Table 3** The correlation values between LH-index and other five centrality measures in four real world networks. The five centrality measures include degree centrality (DC), betweenness centrality (BC), closeness centrality (CC), coreness centrality (KC), and h-index centrality (H-index). The values are obtained by averaging over 100 implementations of the stochastic SIR model.

| Network | DC | BC | CC | KC | H-index |
|---|---|---|---|---|---|
| USair | 0.7922 | 0.5783 | 0.8142 | 0.8250 | 0.8253 |
| Blogs | 0.7766 | 0.5579 | 0.4549 | 0.7518 | 0.7722 |
| Email | 0.8768 | 0.6818 | 0.7932 | 0.8812 | 0.9095 |
| Powergrid | 0.7833 | 0.5814 | 0.2244 | 0.6985 | 0.7997 |

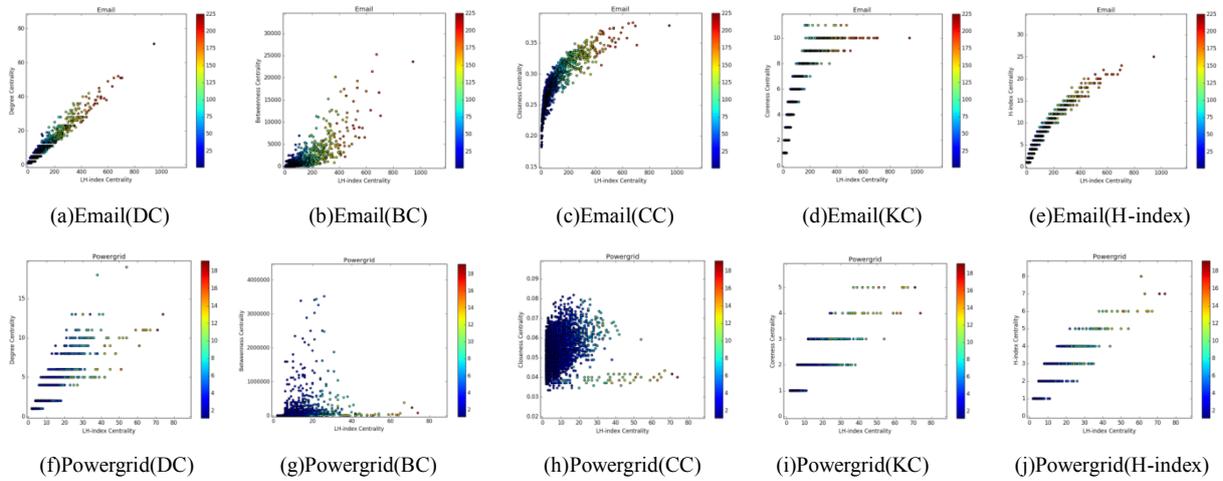

(a)Email(DC)  (b)Email(BC)  (c)Email(CC)  (d)Email(KC)  (e)Email(H-index)

(f)Powergrid(DC)  (g)Powergrid(BC)  (h)Powergrid(CC)  (i)Powergrid(KC)  (j)Powergrid(H-index)

**Fig. 4.** The correlations between LH-index centrality and other five centrality measures in Email and Powergrid. The five centrality measures include degree centrality (DC), betweenness centrality (BC), closeness centrality (CC), coreness centrality (KC), and h-index centrality (H-index). The color represents the number of nodes infected by the initially infected node. The transmission rate is 0.1 for Email and 0.3 for Powergrid according to their epidemic thresholds. The values are obtained by averaging over 100 implementations of the stochastic SIR model.

As can be seen from Fig. 4, for Email, LH-index is positively correlated with DC, CC, KC and H-index. That is to say, with the growth of LH-index values of nodes, the corresponding values of DC, CC, KC and H-index also increase. At the same time, the infected number of nodes also increases obviously with the improvement of nodes' spreading ability. Moreover, LH-index is a little more strongly correlated with KC and H-index compared to DC and CC, which can be found in Table 3. For Email, we can see from Table 3 that the correlation value between LH-index and H-index is 0.9095, which is the highest one among all correlation values. For Powergrid, LH-index is more positively correlated with H-index and DC compared to KC and others, and the corresponding correlation strengths can be found in Table 3. Besides, we also find that LH-index shows its superior resolution power in sparse networks. Take the Powergrid network as an example, we can see from Fig. 4(f) that with the increasing of LH-index values, colors of nodes change from blue to red, but the DC values of these nodes still keep the same, which illustrates that DC cannot distinguish real spreading ability of these nodes effectively.

Overall speaking, LH-index shows its strong positive correlations with H-index, KC and DC. Meanwhile, it is also positively correlated with CC in some dense networks. Therefore, LH-index is a well-balanced mix of these measures, and can own many advantages which originally belong to these centrality measures separately. Besides, LH-index also presents its better resolution power, and thus performs better in identifying nodes' real spreading ability compared to these widely used centrality measures.

**4.4 The performance of different centrality measures in simulated networks**

For better validate the performance of LH-index, we also compare it with other centrality measures in simulated networks generated by two kinds of network models: the Barabasi-Albert (BA) [32] network model and the Lancichinetti-Fortunato-Radicchi (LFR) network model [33].

The Barabasi-Albert (BA) model generates a random scale-free network with the preferential attachment mechanism. The main steps for the BA model to generate a network are listed as follows: Firstly, the network begins with an initial connected network of $m_0$ nodes. Then, each new node is added to the network at a time, and constructs new links with $m < m_0$ nodes with a probability that is proportional to the number of links of existing nodes. Initially, we set the number of nodes $n = 1000$, $m_0 = 20$, and $m = 3, 5$ respectively. Because coreness centrality assigns each connected node with the same value being equal to $m$, and thus loses its detection ability for identifying influential nodes in BA networks, we only compare LH-index with other four centrality measures. The results are presented in Fig. 5.

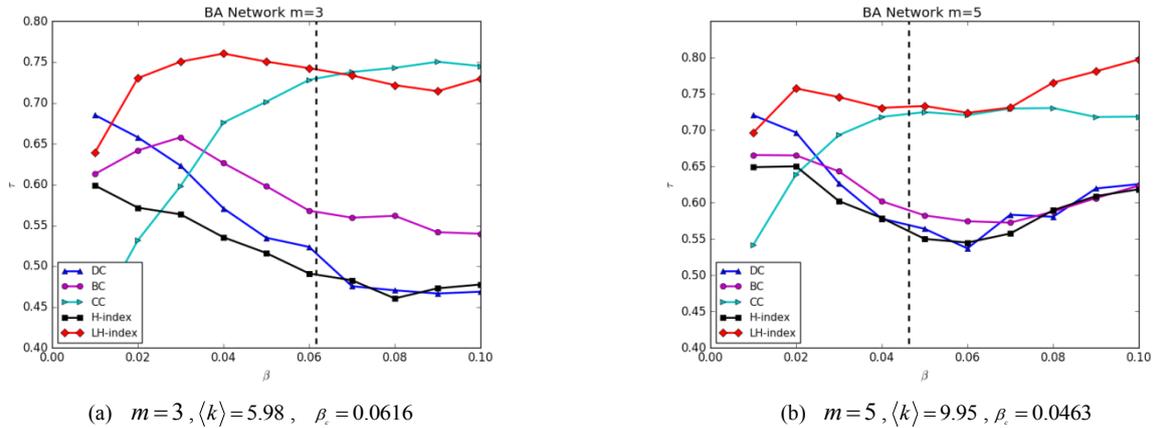

(a) $m = 3$, $\langle k \rangle = 5.98$, $\beta_c = 0.0616$  (b) $m = 5$, $\langle k \rangle = 9.95$, $\beta_c = 0.0463$

**Fig. 5.** The $\tau$ values of DC, BC, CC, H-index and LH-index are obtained by averaging over 500 implementations of the stochastic SIR model under different $\beta$ in BA networks. The vertical dashed line for each plot corresponds to the epidemic threshold $\beta_c$.

As can be seen from Fig. 5, LH-index can also get better performance in a wide range of $\beta$ in BA networks. When $\beta$ is small, DC is better than LH-index, which is the same as previous experiments in real world networks. Due to the characteristics of the BA networks: few nodes unusually own high degree values compared to others in the network. Then, these few nodes, named hubs, are always surrounded by nodes with low degree values. In this kind of network, the performance of H-index method is not satisfactory compared to others, that is because the imbalance of degree distribution leads to most of nodes with low h-index values. However, CC sometimes is superior to LH-index, such as Fig. 5(a), that is mainly because CC usually chooses nodes being close to all other nodes, and the average shortest distance for the BA networks is usually very small, so the hubs are easily selected as influential nodes. Then, when the $\beta$ is larger, it can easily influence a large number of nodes.

The LFR network model is utilized to generate networks with community structure, which is widely used to evaluate the effectivity of one community detection method. Here, we will employ this model to generate networks with different community structures. Then, we check the performance of LH-index in these networks and compare it with other centrality measures. The mixing parameter $\mu$ in LFR is used for tuning the modularity of the generated network. With the increasing of $\mu$, the modularity of the network will decrease, which means the community structure of the generated network will become more inconspicuous.

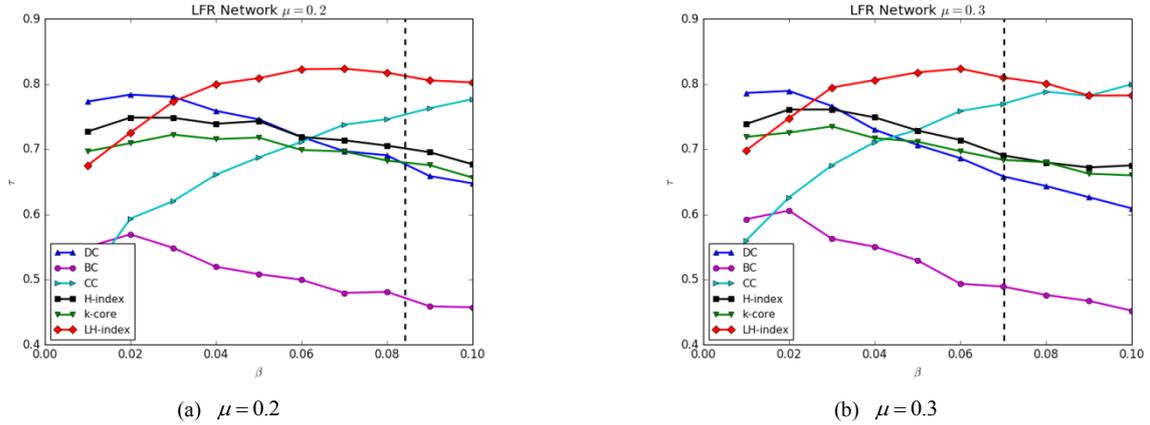

(a) $\mu = 0.2$          (b) $\mu = 0.3$

**Fig. 6.** The $\tau$ values of DC, BC, CC, KC, H-index and LH-index are obtained by averaging over 500 implementations of the stochastic SIR model under different $\beta$ in LFR networks. The vertical dashed line for each plot corresponds to the epidemic threshold $\beta_c$.

    Nematzadeh et al. [43] had investigated the impact of community structure on information spreading, and pointed out that large modularity (smaller $\mu$) facilitates the spreading in the originally community, and small (larger $\mu$) modularity benefits intercommunity spreading. Then, they demonstrated that there exists optimal network modularity for global spreading for a network. According to their research results, we set mix parameter $\mu=0.2$ and $\mu=0.3$, respectively, so as to investigate the performance of LH-index and other measures in LFR networks. Besides, other parameters are set as follows: the average degree of nodes $\langle k \rangle = 5$, the maximum degree of nodes $k_{max}=50$, the exponent of degree distribution is 2, the exponent of community size distribution is 1, the community size range is [20,50]. The clustering coefficients of these two networks are 0.3430 and 0.2687, respectively. The results are presented in Fig. 6. We can see that LH-index presents its better performance in measuring nodes' influences in networks with community structure compared to other methods. Communities in a network are usually connected by some edges, and nodes located in the periphery of communities, named border nodes, are connected by these edges. Obviously, the border nodes are important for the influence diffusion between communities. Therefore, how to evaluate the spreading ability of these nodes is important. DC only considers the number of neighbors, but the degree values of border nodes are always small, thus these nodes are rarely selected as information spreaders. For instance, in Fig. 6(a), the performance of DC decreases quickly when $\beta$ grows larger. In Fig. 6, we can see that H-index is better than DC when $\beta$ is close to $\beta_c$. That is because H-index takes into account of neighbors' degree information of a border node, and then can easily choose the border node with high influence. However, H-index assigns too many border nodes with the same value, which is difficult for us to distinguish these nodes' real spreading ability, so this condition leads to the declining of its performance. For LH-index, it utilizes neighbor nodes' h-index information of a border node, which can better measure the influence of this node. Then, LH-index can evaluate border nodes' spreading ability across communities more effectively, and can achieve better performance in a wide range of $\beta$. The experimental results presented in Fig. 6(a) and Fig. 6(b) can well illustrate this point.

    According to the performance of LH-index in BA networks and LFR networks with community structure, we can also demonstrate the superiority of LH-index in identifying and ranking influential nodes in simulated networks.

## 3   Conclusions

    In this paper, we propose a novel centrality measure, LH-index centrality, to better identify and rank influential spreaders in networks. The basic idea of LH-index is that one node connecting to more influential nodes will also be much more influential. Then, we utilize the stochastic SIR model in several networks to evaluate the performance of LH-index method in identifying influential nodes. In addition, we also compare the performance of LH-index with other

centrality methods, such as DC, BC, CC, KC and H-index centrality.

For future works, we will pay more attention to the variation of transmission rate on the performance of centrality measures, which has not been studied thoroughly until now. Meanwhile, considering that most real networks are dynamic, so we will also concentrate on designing effective centrality methods for dynamic networks.

## Acknowledgements

We thank the editor and the three anonymous reviewers for their insightful and constructive comments. This work was supported by 973 Program of China (Grant Nos.2013CB329601, 2013CB329604, and 2013CB329606) and National Defense Science and Technology Project Foundation of China (Grant No.3101283).

## References

[1] C. Wang, W. Chen, Y. Wang, Scalable influence maximization for independent cascade model in large-scale social networks, Data Min. Knowl. Discov. 25 (2012) 545-576.

[2] J. Zhang, J. Tang, J. Li, Y. Liu, C. Xing, Who influenced you? Predicting retweet via social influence locality, ACM Trans. Knowl. Data Eng. D 9 (2015) 25.

[3] W. Xu, Z. Lu, W. Wu, Z. Chen, A novel approach to online social influence maximization, Soc. Netw. Anal. Min. 4 (2014) 153.

[4] W. Chen, C. Wang, Y. Wang, Scalable influence maximization for prevalent viral marketing in large-scale social networks, in: Proceedings of the 16th ACM SIGKDD International Conference on Knowledge Discovery and Data Mining, 2010, pp. 1029-1038

[5] H. Zhang, J. Zhang, C. Zhou, M. Small, B. Wang, Hub nodes inhibit the outbreak of epidemic under voluntary vaccination, New J. Phys. 12 (2010) 023015.

[6] C. Nowzari, V.M. Preciado, G.J. Pappas, Analysis and control of epidemics: A survey of spreading processes on complex networks, IEEE Control Syst. 36 (2016) 26-46.

[7] Z. Lin, X. Zheng, N. Xin, D. Chen, CK-LPA: Efficient community detection algorithm based on label propagation with community kernel, Physica A 416 (2014) 386-399.

[8] H. Sun, J. Liu, J. Huang, G. Wang, Z. Yang, Q. Song, X. Jia, CenLP: A centrality-based label propagation algorithm for community detection in networks, Physica A 436 (2015) 767-780.

[9] A. Czaplicka, J.A. Holyst, P.M.A. Sloot, Noise enhances information transfer in hierarchical networks, Sci. Rep. 3 (2013) 1223.

[10] G. Lawyer, Understanding the influence of all nodes in a network, Sci. Rep. 5 (2015) 8665.

[11] L.C. Freeman, Centrality in social networks conceptual clarification, Social Networks 1 (1978) 215-239.

[12] L.C. Freeman, A set of Measures of centrality based on betweenness, Sociometry, 40 (1977) 35-41.

[13] G. Sabidussi, The centrality index of a graph, Psychometrika 31 (1966) 581-603.

[14] M. Kitsak, L.K. Gallos, S. Havlin, F. Liljeros, L. Muchnik, H.E. Stanley, H.A. Makse, Identification of influential spreaders in complex networks, Nat. Phys. 6 (2010) 888-893.

[15] L. Lü, Y.-C. Zhang, C.H. Yeung, T. Zhou, Leaders in social networks, the delicious case, PLoS One 6 (2011) e21202.

[16] D. Chen, L. Lü, M.-S. Shang, Y.-C. Zhang, T. Zhou, Identifying influential nodes in complex networks, Physica A 391 (2012) 1777-1787.

[17] J. Bae, S. Kim, Identifying and ranking influential spreaders in complex networks by neighborhood coreness, Physica A 395 (2014) 549-559.

[18] L.-L. Ma, C. Ma, H.-F. Zhang, B.-H. Wang, Identifying influential spreaders in complex networks based on gravity formula, Physica A 451 (2016) 205-212.

[19] Y. Liu, M. Tang, T. Zhou, Y. Do, Identify influential spreaders in complex networks, the role of neighborhood, Physica A 452 (2016) 289-298.

[20] F.D. Malliaros, M.-E.G. Rossi, M. Vazirgiannis, Locating influential nodes in complex networks, Sci. Rep. 6 (2016) 19307.

[21] S. Basu, U. Maulik, A game theory inspired approach to stable core decomposition on weighted networks, IEEE Trans. Knowl. Data Eng. 28 (2016) 1105-1117.

[22] Y. Liu, M. Tang, T. Zhou, Y. Do, Core-like groups result in invalidation of identifying super-spreader by k-shell decomposition, Sci. Rep. 5 (2015) 09602.

[23] Y. Liu, M. Tang, T. Zhou, Y. Do, Improving the accuracy of the k-shell method by removing redundant links: From a perspective of spreading dynamics, Sci. Rep. 5 (2015) 13172.

[24] J.E. Hirsch, An index to quantify an individual's scientific research output, Proc. Natl. Acad. Sci. USA 102 (2005) 16569-16572.

[25] A. Korn, A. Schubert, A. Telcs, Lobby index in networks, Physica A 388 (2009) 2221-2226.

[26] L. Lü, T. Zhou, Q.-M. Zhang, H.E. Stanley, The H-index of a network node and its relation to degree and coreness, Nature Commun. 7 (2016) 10168.

[27] L. Lü, D. Chen, X.-L. Ren, Q.-M. Zhang, Y.-C. Zhang, T. Zhou, Vital nodes identification in complex networks, Phys. Rep. 650 (2016) 1-63.

[28] S.N. Dorogovtsev, A.V. Goltsev, J.F.F. Mendes, k-Core organization of complex networks, Phys. Rev. Lett. 96 (2006) 040601.

[29] V. Batagelj, A. Mrvar, Pajek data sets, (2003) Available: http://vladowiki.fmf.uni-lj.si/doku.php?id=pajek:data:pajek:vlado.

[30] R. Guimera, L. Danon, A. Diaz-Guilera, F. Giralt, A. Arenas, Self-similar community structure in a network of human interactions, Phys. Rev. E 68 (2003) 065103.

[31] D.J. Watts, S.H. Strogatz, Collective dynamics of 'small-world' networks, Nature 393 (1998) 440-442.

[32] A.-L. Barabási, R. Albert, Emergence of scaling in random networks, Science 286 (1999) 509-512.

[33] A. Lancichinetti, S. Fortunato, F. Radicchi, Benchmark graphs for testing community detection algorithms, Phys. Rev. E 78 (2008) 046110.

[34] C. Castellano, R. Pastor-Satorras, Thresholds for epidemic spreading in networks, Phys. Rev. Lett. 105 (2010) 218701.

[35] W.O. Kermack, A.G. McKendrick, A contribution to the mathematical theory of epidemics, Proc. R. Soc. Lond. Ser. A Math. Phys. Eng. Sci. 115 (1927) 700–721.

[36] Q. Ma, J. Ma, Identifying and ranking influential spreaders in complex networks with consideration of spreading probability, Physica A 465 (2017) 312-330.

[37] S. Gao, J. Ma, Z. Chen, G. Wang, C. Xing, Ranking the spreading ability of nodes in complex networks based on local structure, Physica A 403 (2014) 130-147.

[38] A. Garas, P. Argyrakis, C. Rozenblat, M. Tomassini, S. Havlin, Worldwide spreading of economic crisis, New J. Phys. 12 (2010) 113043.

[39] H.C. Tuckwell, R.J. Williams, Some properties of a simple stochastic epidemic model of SIR type, Math. Biosci. 208 (2007) 76-97.

[40] D.T. Gillespie, Exact stochastic simulation of coupled chemical reactions, J. Phys. Chem. 81 (1977) 2340-2361.

[41] F. Radicchi, C. Castellano, Leveraging percolation theory to single out influential spreaders in networks, Phys. Rev. E 93 (2016) 062314.

[42] M.G. Kendall, A new measure of rank correlation, Biometrika 30 (1938) 81-93.

[43] A. Nematzadeh, E. Ferrara, A. Flammini, Y.Y. Ahn, Optimal network modularity for information diffusion, Phys. Rev. Lett. 113 (2014) 088701.